\documentclass[preprint,12pt]{elsarticle}

\usepackage{epsfig}

\usepackage{amssymb}
\journal{Physica B}

\begin{document}

\begin{frontmatter}



\title{A dynamical mean-field study of orbital-selective Mott phase
enhanced by next-nearest neighbor hopping}


\author{Yuekun Niu$^{1}$}

\author{Jian Sun$^{1, 2}$}

\author{Yu Ni$^{1}$}

\author{Yun Song$^{\dag 1} $}
\ead{yunsong@bnu.edu.cn}

\address{$^1$Department of Physics, Beijing Normal University,
Beijing 100875, China}

\address{$^2$Beijing National Laboratory for Condensed Matter Physics,
Institute of Physics, Chinese Academy of Sciences, Beijing 100190, China}

\begin{abstract}
The dynamical mean-field theory is employed to study the
orbital-selective Mott transition (OSMT) of the two-orbital
Hubbard model with nearest neighbor hopping and
next-nearest neighbor (NNN) hopping. The NNN hopping breaks
the particle-hole symmetry at half filling and gives rise to
an asymmetric density of states (DOS). Our calculations show
that the broken symmetry of DOS benefits the OSMT, where
the region of the orbital-selective Mott phase significantly
extends with the increasing NNN hopping integral. We also
find that Hund's rule coupling promotes OSMT by blocking the
orbital fluctuations, but the influence of NNN hopping is
more remarkable.
\end{abstract}

\begin{keyword}


Dynamical mean-field theory \sep Two-orbital Hubbard model \sep
Orbital-selective Mott transition \sep Next-nearest neighbor hopping

\PACS 71.27.+a \sep 71.30.+h \sep 72.80.Ga

%71.27.+a   Strongly correlated electron systems; heavy fermions
%71.30.+h   Insulator-metal transitions and other electronic transitions
%72.80.Ga   Transition-metal compounds


\end{keyword}
\end{frontmatter}



\section{Introduction}
\label{Introduction}

Mott metal-insulator transitions (MIT) in strongly correlated electron
systems with orbital degrees of freedom has received much attention
over the past decades \cite{Kotliar-2006, Georges-2013}.
The orbital fluctuations tuned by the interactions may force the Mott
MIT to happen successively in different orbitals, leading to the
so-called orbital-selective Mott transition (OSMT) in the
non-degenerated multiorbital systems \cite{Georges-2013, Anisimov-2002}.
In an orbital-selective Mott phase (OSMP), the carriers on a subset of
orbitals get localized but others remain itinerant. This phenomenon has
been observed experimentally in some transition-mental compounds,
including the iron-based superconductors
\cite{Arcon-2010, Miao-2016}.

Many theoretical methods have been employed to study the OSMT in
multiorbital systems, including the quantum Monte Carlo technique
\cite{Bouadim-2009}, dynamical cluster approximation \cite{LeeH-2010},
slave-boson method \cite{Kotliar-1986}, and dynamical mean-field
theory (DMFT)\cite{Kotliar-2006, Gull-2011}. It is well known that the
DMFT approach, which handles band-like and atomic-like aspects on equal
footing, is an appropriate theoretical framework for the study of Mott
MIT \cite{Georges-1996}. Combined with various impurity solvers, the
DMFT approach has been used to study the OSMT in multiorbital systems
with different energy scales. So far, three kinds of factors have
been confirmed for the appearance of  OSMT, including the Hund's rule
coupling, crystal field splitting, and bandwidth differences
among orbitals.

It has been proposed that Hund's rule coupling ($J$) is indeed
responsible for the correlation effects by strongly suppressing
the coherence scale for the formation of a Fermi liquid
\cite{Georges-2013}. The Hund's coupling promotes the OSMT at
half filling \cite{Sun-2015, deMedici-2011, Liebsch-2005},
which can be understood by recognizing that $J$ blocks orbital
fluctuations \cite{Georges-2013}. Besides, the other two factors
influence the OSMT by introducing the nondegeneration among the
orbitals of multiorbitals system. The important role of the bandwidth
difference in finding OSMT has been verified by the earlier DMFT
investigations of the two-orbital Hubbard model with unequal bandwidths
\cite{Koga-200405, deMedici-2005, Song-2005}.
On the other hand, the orbital degeneracy can also be broken by the
crystal field splitting, leading to OSMT  in multiorbital systems
\cite{Jakobi-2013, Song-2009,  deMedici-2009, Werner-2007}.

Apart from the three factors mentioned above, is there any other
model parameter which also plays an essential role in the OSMT? In this
study we concentrate on the influence of the next-nearest neighbor
(NNN) hopping integrals. In some previous DMFT calculations
\cite{Georges-1996}, the randomness of the NNN hopping has been
introduced to suppress antiferromagnetism in the half-filled Hubbard
model at weak coupling. Under this condition, the density of states (DOS)
remains semielliptic, and the Bethe lattice holds the particle-hole symmetry
\cite{Rozenberg-1995,Zitzler-2004}. However, it has been found that
the DOS is no longer semielliptic in the tight-binding model with
standard NNN hopping \cite{Eckstein-2005,Peters-2009}.
Therefore, it is important to make it clear how the OSMT is influenced by
the broken symmetry of DOS for multiorbital systems at half-filling.

In this paper we study the effect of NNN hopping on the OSMT in
two-orbital Hubbard model by using the DMFT approach with the Lanczos
diagonalization method \cite{Dagotto-1994} as its impurity solver.
The Lanczos solver is very powerful in finding the critical point of the
Mott MIT, which has been proved to be far superior than some other
impurity solvers \cite{Caffarel-1994, Chitra-1999}.
Because of the asymmetric DOS introduced by the NNN hopping, the
calculations become more complicated. We have to adjust the value of
chemical potential to construct the whole phase diagram of the
two-orbital Hubbard model at half filling. By performing a large amount
of numerical calculations, we investigate the evolution of phase diagram
with the increasing of the NNN hopping integrals in the conditions with
different Hund's rule coupling and also different nearest neighbor
hopping ratio. We find that the region of OSMP significantly increases
with the increasing NNN hopping amplitudes, indicating that the asymmetric
DOS introduced by the NNN hopping plays a key role in the promotion of
OSMT, in spite of the change of the bandwidth ratio. On the other hand,
we also find that the orbital fluctuations are blocked with the increasing
of Hund's rule coupling, leading the critical values of both narrow and wide
band decrease manifestly. However, the enhancement of Hund's coupling
on the OSMT is weaker than the effect of the NNN hopping.

\section{Model and methodology}
\label{Methodology}

We study the extended two-orbital Hubbard model, where the hopping has both
NN and NNN contributions. The Hamiltonian is expressed as
\begin{eqnarray}
    H&=&-\sum_{l}t_{l}\sum_{< ij >\sigma}
              d^{\dag}_{il\sigma} d_{jl\sigma}
        -\sum_{l}t^{\prime}_{l}\sum_{\ll ii^{\prime}\gg\sigma}
              d^{\dag}_{il\sigma} d_{i^{\prime}l\sigma}
        -\mu\sum_{il\sigma}d^{\dag}_{il\sigma} d_{il\sigma}
        \nonumber\\
      &&+\frac{U}{2}\sum_{il\sigma}
              n_{il\sigma} n_{il\bar{\sigma}}
        +\sum_{i\sigma\sigma'}(U'-\delta_{\sigma\sigma'}J)
              n_{i1\sigma}n_{i2\sigma'}
         \nonumber\\
      &&+\frac{J}{2}\sum_{i,l\neq l',\sigma}(
              d^{\dag}_{il\sigma} d^{\dag}_{il\bar{\sigma}}
              d_{il'\bar{\sigma}} d_{il'\sigma}
             +d^{\dag}_{il\sigma} d^{\dag}_{il'\sigma'}
              d_{il\sigma'} d_{il'\sigma}),
\label{Hub-J}
\end{eqnarray}
where operator $d^{\dag}_{il\sigma}$ creates an electron with spin $\sigma$
in orbital $l$ of site $i$.  $< ij >$ and $\ll ii^{\prime}\gg$ represent
the summations over NN and NNN sites, and $t_l$ and $t^{\prime}_l$ denote
the NN and NNN hopping amplitudes for orbital $l$. $U$ and
$U^{\prime}$ are the intra-orbital and interorbital Coulomb interactions,
and $J$ is the Hund's rule coupling.
The onsite component of Green's function for
different orbital $l$ can be obtained by \cite{Georges-1996},
\begin{equation}
G^{(l)}_{ii}(\omega)=\sum_{\vec{k}} G_{l}(\vec{k},\omega)=\int^{+\infty}_{-\infty} d\epsilon \frac{D_{l}^{t, t^{\prime}}(\epsilon)} {\omega+\mu-\epsilon_{l}(\vec{k})-\Sigma_{l}(\omega)}.
\label{GFunc}
\end{equation}
In the infinite limit $Z\rightarrow \infty $, the DOS of the Bethe lattice
with both NN and NNN hopping can be expressed as \cite{Eckstein-2005}
\begin{equation}
D_{l}^{t,t^{\prime}}(\epsilon)=\frac{\Theta[1+4K_{l}(K_{l}+\epsilon/t_{l})]}
{\sqrt{1+4K_{l}(K_{l}+\epsilon/t_{l})}}
\sum_{n=1}^{2}\frac{\sqrt{4-[\lambda_{l}^{(n)}]^2(\epsilon)}}{2\pi t_{l}},
\end{equation}
with
\begin{eqnarray}
\lambda_{l}^{(1)}(\epsilon)=\frac{-1 +\sqrt{1+4K_{l}(K_{l}+\epsilon/t_{l})}}{2K_{l}},
\nonumber\\
\lambda_{l}^{(2)}(\epsilon)=\frac{-1 -\sqrt{1+4K_{l}(K_{l}+\epsilon/t_{l})}}{2K_{l}},
\end{eqnarray}
where $K_{l}=t_{l}^{\prime}/t_{l}$ represents the ratio between NNN hopping $t_{l}^{\prime}$ and NN hoping $t_l$ of orbital $l$. As the NNN hopping integrals
increase, $D_{l}^{t,t^{\prime}}(\epsilon)$ becomes asymmetric and develops
a square-root singularity at a band edge \cite{Eckstein-2005}.

In the framework of DMFT, the Hubbard model is mapped into an Anderson impurity
model (AIM),
\begin{eqnarray}
    H_{imp}&=&\sum_{ml\sigma}\epsilon_{ml} c^{\dag}_{ml\sigma}c_{ml\sigma} +\sum_{ml\sigma}V_{ml}(c^{\dag}_{ml\sigma}d_{l\sigma}
    +d^{\dag}_{l\sigma}c_{ml\sigma})
    +\sum_{l\sigma}(\epsilon_l-\mu)d^{\dag}_{l\sigma}d_{l\sigma}\nonumber\\
    &+&\frac{U}{2}\sum_{l\sigma}n_{l\sigma}n_{l\bar{\sigma}}+\sum_{\sigma\sigma'}
    (U'-\delta_{\sigma\sigma'}J)n_{1\sigma}n_{2\sigma'}
    +\frac{J}{2}\sum_{l\neq l',\sigma}d^{\dag}_{l\sigma}d^{\dag}_{l\bar{\sigma}}d_{l'\bar{\sigma}}d_{l'\sigma}\\
    &+&\frac{J}{2}
    \sum_{l\neq l',\sigma}d^{\dag}_{l\sigma}d^{\dag}_{l'\sigma'}d_{l\sigma'}d_{l'\sigma},\nonumber\label{IMP}
\end{eqnarray}
where the parameter $\epsilon_{ml}$ represents the energy of the $m$th
environmental bath for the orbital $l$, and $V_{ml}$ describes the
couplings between the bathes and the impurity site.

We employ Lanczos exact diagonalization approach \cite{Dagotto-1994} as
an impurity solver to calculate the Green's function ($G_{AIM}^{(l)}$)
and the self energy ($\Sigma_{AIM}^{(l)}$) of AIM.
The parameters $\epsilon_{ml}$ and $V_{ml}$ in AIM
can be obtained self-consistently by intorducing
$G_{ii}^{(l)}(\omega)=G_{AIM}^{(l)}(\omega)$ and
$\Sigma_l(\omega)=\Sigma_{AIM}^{(l)}(\omega)$ \cite{Georges-1996}.
In our DMFT calculations, the bath size is chosen as $n_{b}=3$.
In Table~1, we show the self-consistent values of the parameters of
AIM for the metallic phase, OSMP and insulating phase, respectively.

\begin{table*}[htb]
\begin{center}
\caption{The values of the AIM parameters in the DMFT self-consistent
calculations for different interactions $U$ when $t_2/t_1=0.5$,
$K=0.5$, and $J/U=0.5$, corresponding to the metallic phase,
orbital-selective Mott phase and insulating phase, respectively.} \
%{\scriptsize %%%%%%%%%%%%%%%%%%%%
\begin{tabular}{c|rrrrr}
\hline \hline \\ [-4pt]
  Metal $(U=0.01)$
%\\ [+2pt]
%\hline \\ [-4pt]
%  $\backslash$ $ |\bm{R_i}-\bm{R_j}|$
% $(\alpha, \beta)$ $\backslash$ $|\bm{R_i}-\bm{R_j}|$
 & bath-1
 & bath-2
 & bath-3
\\ [+4pt]
 \hline \\ [-5pt]             %  000     1/2-1/20  100     1-10   3/2-1/20 00c/a  1/2-1/2c/a sigy I sigd
 $\epsilon_{1}$             &            1.498717 &            0.334170 &          -0.072524 &                 \\ [+4pt]%%
 $V_{1}$                    &            0.862313 &            0.411842 &           0.423400 &                 \\ [+4pt]%%
 $\epsilon_{2}$             &            0.376248 &           -0.108999 &           0.005459 &                 \\ [+4pt]%%
 $V_{2}$                    &            0.425066 &            0.171245 &          -0.109655 &                 \\ %%
\hline \\ [-4pt]
% $(\alpha, \beta)$ $\backslash$ $|\bm{R}_i-\bm{R}_j|$
  OSMP $(U=2.40)$
 & bath-1
 & bath-2
 & bath-3
\\ [+4pt]
 \hline \\ [-5pt]             %  000     1/2-1/20  100     1-10   3/2-1/20 00c/a  1/2-1/2c/a sigy I sigd
 $\epsilon_{1}$             &            0.780082 &            0.100286 &          -0.015272 &                 \\ [+4pt]%%
 $V_1$                      &            0.486443 &            0.195582 &           0.178468 &                 \\ [+4pt]%%
 $\epsilon_{2}$             &            1.435533 &           -0.875888 &          -0.280482 &                 \\ [+4pt]%%
 $V_2$                      &            0.338925 &            0.269218 &           0.000639 &                 \\ %%
\hline \\ [-4pt]
% $(\alpha, \beta)$ $\backslash$ $|\bm{R}_i-\bm{R}_j|$
  Insulator ($U=4.40)$
 & bath-1
 & bath-2
 & bath-3
\\ [+4pt]
 \hline \\ [-5pt]             %  000     1/2-1/20  100     1-10   3/2-1/20 00c/a  1/2-1/2c/a sigy I sigd
 $\epsilon_{1}$            &             3.021307 &           -1.608611 &         -0.268511 &                 \\ [+4pt]%%
 $V_1$                     &             0.701389 &            0.518544 &         -0.000135 &                 \\ [+4pt]%%
 $\epsilon_{2}$            &             2.940402 &           -2.100353 &         -1.077004 &                 \\ [+4pt]%%
 $V_2$                     &             0.337240 &            0.285466 &         -0.001323 &                 \\ %%

\hline \hline
\end{tabular}
%} %%%%%%%%%%%%%%%%%%
\end{center}
\label{MP}
\end{table*}

In the next section, a large amount of DMFT calculations are conducted to
construct the whole phase diagram, presenting the effect of NNN hoping on
Mott MIT in two-orbital Hubbard model.

\section{Results}
\label{OSMT}

\begin{figure}[ht]
\centerline{\epsfxsize=5.0in\epsfbox{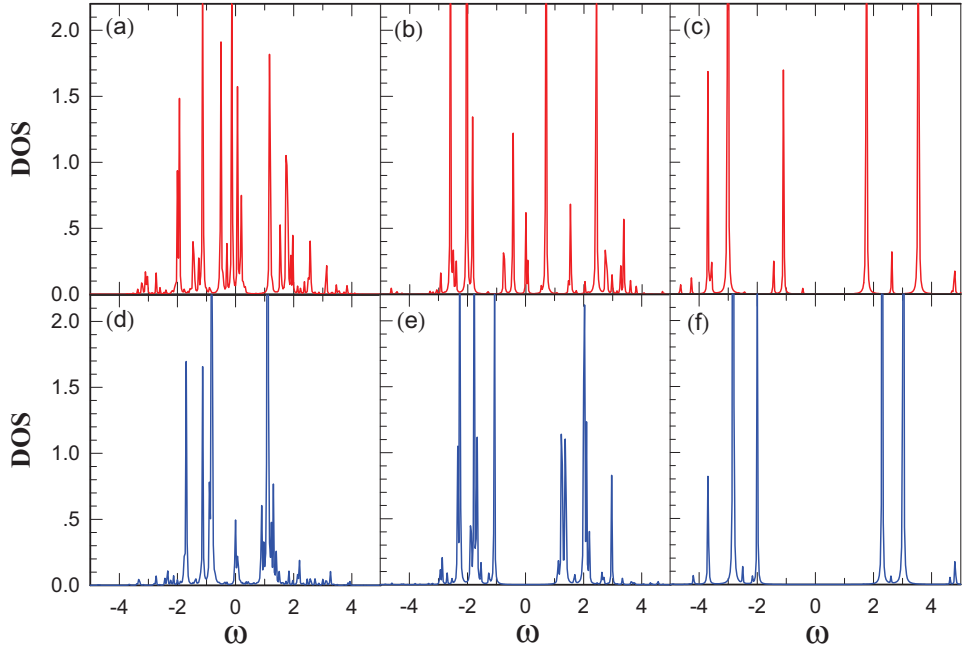}}
\caption{(Color online) Density of states of
  wide band (up\emph{}per panel) and narrow band (lower panel)
  for different onsite interactions:
  $U=2.0$ ((a) and (d)), $U=3.0$ ((b) and (e)),
  and $U=4.5$ ((c) and (f)).
  The parameters of the two-orbital Hubbard model are:
  $t_2/t_1=0.4$, $K=t_1^{\prime}/t_1=t_2^{\prime}/t_2=0.1$,
  $J=U/4$, and $U^{\prime}=U-2J$.
  $t_1$ ($t_2$) and $t_1^{\prime}$ ($t_2^{\prime}$) are
  nearest neighbor and next-nearest neighbor hopping
  integrals of the wide (narrow) band. Energies are in unit $t_1$,
  and the energy broadening factor is $\epsilon$=0.01.}
\label{fig1}
\end{figure}

The DOS of standard Hubbard model is particle-hole symmetric at
half filling. Nevertheless, in future consideration of the NNN
hopping integrals, the DOS becomes asymmetric at half filling.
Fig.~\ref{fig1} shows the DOS of the two-orbital Hubbard model with
$K=t_1^{\prime}/t_1=t_2^{\prime}/t_2=0.1$ for different interactions $U$,
where $t_1$ and $t_2$ are the NN hopping for the wide and narrow bands,
and $t_1^{\prime}$ and $t_2^{\prime}$ represent the NNN hopping
accordingly. The particle-hole symmetry of the DOS is broken by the
NNN hopping for both the wide and narrow bands, and the asymmetry
becomes more distinct in the conditions with weak interactions.

\begin{figure}[ht]
\centerline{\epsfxsize=5.0in\epsfbox{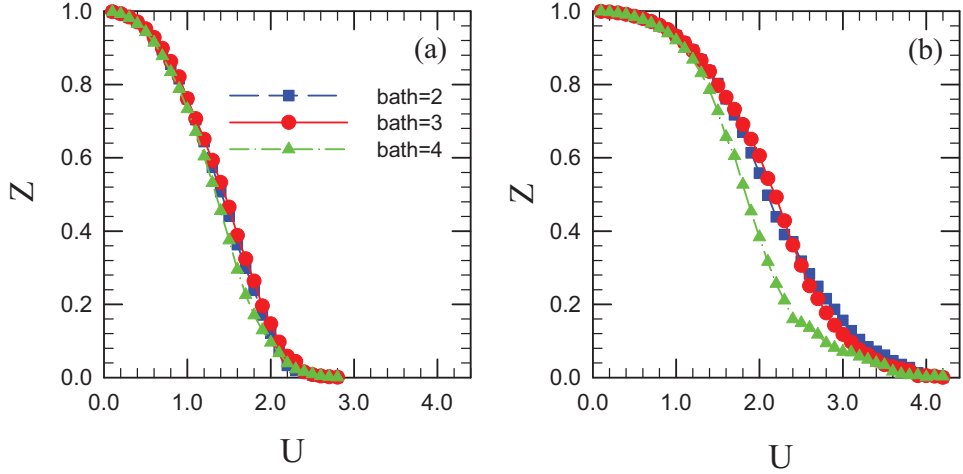}}
\caption{(Color online) Interaction dependence of
  the quasiparticle weight $Z$ for the narrow band (a)
  and wide band (b) when $t_2/t_1=0.4$,
  $K=t_1^{\prime}/t_1=t_2^{\prime}/t_2=0.7$,
  $J=U/4$, and $U^{\prime}=U-2J$.
  The results for different bath numbers $n_b=2$, 3 and 4
  in the DMFT calculations are shown by the squares,
  circles and triangles, respectively.
  Energies are in unit $t_1$.}
\label{fig2}
\end{figure}

Most of the theoretical studies have paid attention to the OSMT in
the systems with particle-hole symmetry at half filling
\cite{Koga-200405,deMedici-2005,Song-2005,Tocchio-2016}.
By introducing the NNN hopping, we could find out whether the OSMT
exists when the particle-hole symmetry is broken at half filling.
As shown in Fig.~\ref{fig1}(a) and (d), resonance peaks appear at Fermi
level of DOS of both the wide and narrow bands for weak interactions
$U=2.0$, suggesting that the two orbitals are all metallic.
OSMP appears when the onsite interaction increases to $U=3.0$, as shown
in Fig.~\ref{fig1}(b) and \ref{fig1}(e). In Fig.~\ref{fig1}(b) the wide
band is still metallic with resonance peaks at Fermi level, but a Mott
gap opens around the Fermi level of the narrow band (Fig.~\ref{fig1}(e)).
Further increasing interactions to $U=4.5$, both the wide and narrow
bands transit to Mott insulating phase. Therefore, OSMT is still found
in two-orbital Hubbard model with both NN and NNN hopping, where the DOS
is asymmetric at half filling.

\begin{figure}[ht]
\centerline{\epsfxsize=4.5in\epsfbox{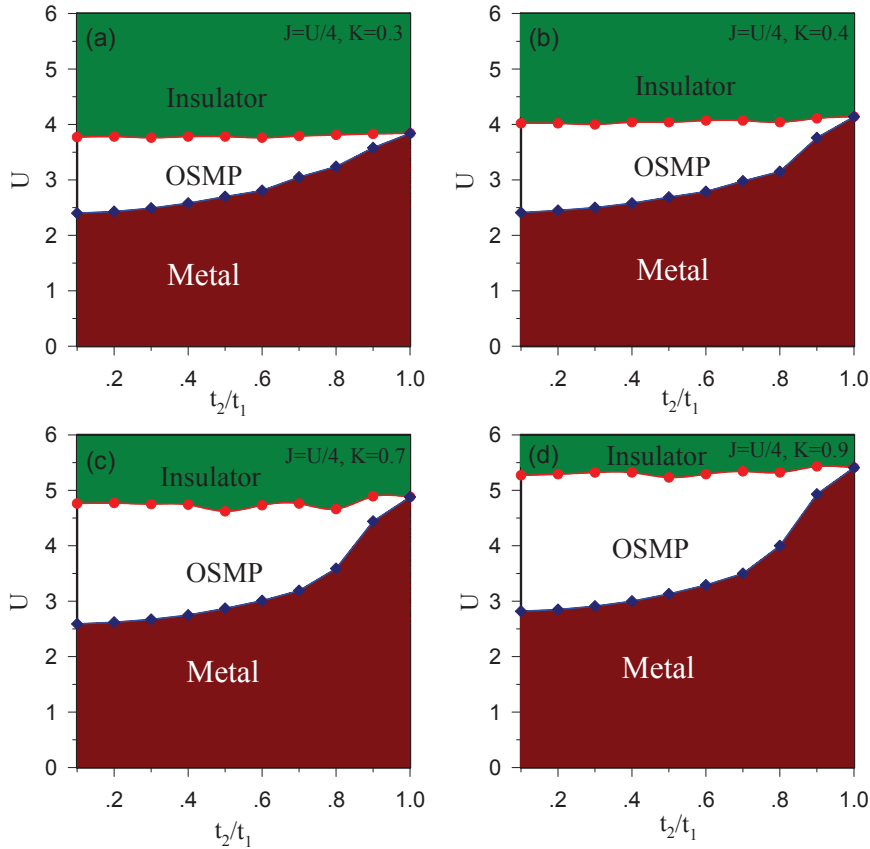}}
\caption{(Color online) Effect of the ratio between
  nearest neighbor hopping $t_2/t_1$ on
  orbital-selective Mott transition for
  different next-nearest neighbor hoping
  ($K=t_1^{\prime}/t_1=t_2^{\prime}/t_2$):
  $K=0.3$ (a), $K=0.4$ (b), $K=0.7$ (c),
  and $K=0.9$ (d). The Hund's rule couplings
  are chosen as $J=U/4$, and the energies are
  in units of $t_1$.}
\label{fig3}
\end{figure}

The critical points of OSMT could be determined precisely by studying
the interaction dependence of the quasiparticle weight of each orbital
($Z_l$), which is defined by
\begin{equation}
Z_l=\{1-\frac{dRe\Sigma_l(\omega)}{d\omega}|_{\omega=0}\}^{-1},
\end{equation}
where $\Sigma_l(\omega)$ represents the self-energy of orbital $l$.
Decreasing with the increasing interactions $U$ as shown in
Fig.~\ref{fig2}, the quasiparticle weights of different orbital drops
to zero successively, indicating the appearance of OSMT. When the
parameters of the two-orbital Hubbard model are $t_2/t_1=0.4$,
$K=0.7$ and $J=U/4$, the Mott transition happens first in the narrow band at
$U_{c2}=2.75$, which is much smaller than the critical value of
the wide band $U_{c1}=4.0$. To show the influence of the bath size
on the critical points, we plot in Fig.~\ref{fig2} the results for
different bath numbers $n_{b}$=2, 3 and 4. We find that, in the two-orbital
Hubbard model, the critical values of the Mott transition obtained by
the cases with different bath size are very close to each other for both the
narrow and wide orbitals. It is worth noting that, as $Z_l$ drops
continuously to zero, the Mott transitions are of second order for both
orbitals.

In order to fully understand the effect of NNN hopping on OSMT,
we have performed a great amount calculations to obtain the phase
diagrams of the two-orbital Hubbard model with different $K$.
As mentioned above, the NNN hopping will introduce a particle-hole
asymmetric DOS. We have to adjust the chemical potential to make
the two orbitals are all half filled. As shown in
Fig.~\ref{fig3}, the NN hopping plays an essential rule for the
appearance of OSMP, where the relationship $t_1\neq t_2$ should
be satisfied. Because the value of the NN hopping of wide band is
kept as $t_1$=1, the critical value $U_{c1}$ for wide band remains
unchanged in all four phase diagrams.
Whereas $U_{c2}$ for the narrow band increases continuously with the
increasing $t_2/t_1$.
Apart from the nondegeneration of the two orbitals resulted from the
unequal of the NN hopping $t_1$ and $t_2$, significant influence of
NNN hopping on the OSMP  has also been observed in Fig.~\ref{fig3}.

On the other hand, the area of OSMP expended with the increase of $K$,
suggesting that the NNN hopping is in favor of the OSMT.
Obviously, the contribution for the extension is mainly from the
elevation of the Mott transition point of the wide band $U_{c1}$.
For example, when $t_2/t_1=0.1$, $U_{c1}$ increase
about $45\%$ as $K$ increases from 0.3 to 0.9. While, the
corresponding change for $U_{c2}$ is only $12.5\%$.
Our finding indicates that the influence of the assymmetric DOS
on the OSMT is stronger in the wide band than in the narrow band.
To understand this phenomena, we should consider the interplay
between the multiorbital correlations and the NNN hopping
rather than the effect of bandwidth ratio.

\begin{figure}[ht]
\centerline{\epsfxsize=4.25in\epsfbox{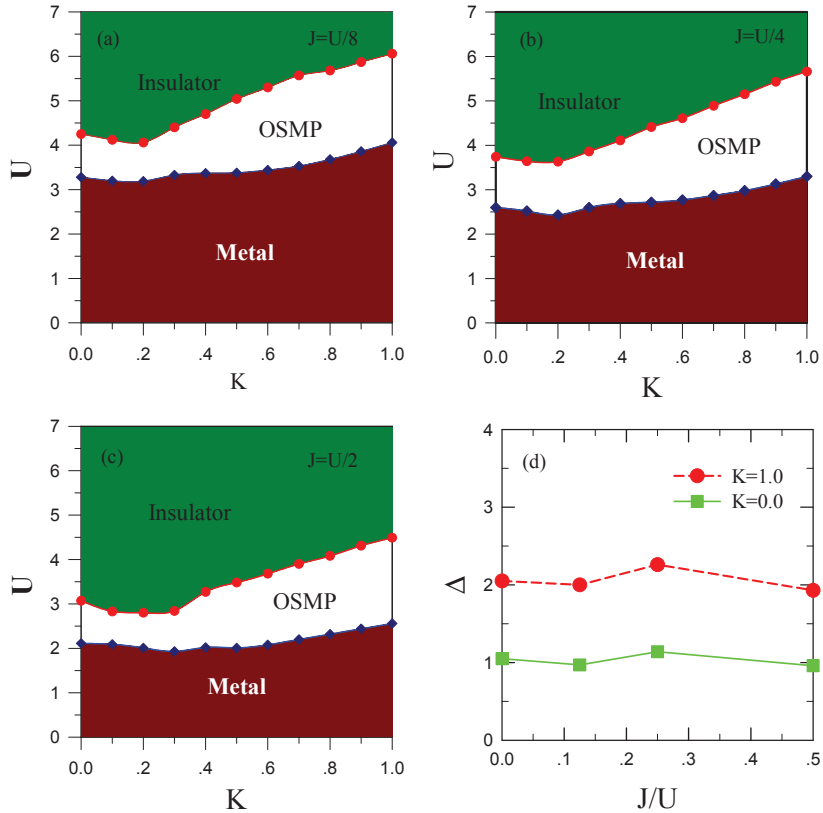}}
\caption{(Color online) Phase diagrams for the systems with
  different Hund's rule   coupling:  (a) $J=U/8$, (b) $J=U/4$,
  and (c) $J=U/2$.
  (d) The $J/U$ dependence of the difference between the
  critical interactions of the wide and narrow band
  ($\Delta=U_{c1}-U_{c2}$) for system with or without
  next-nearest   neighbor hopping.
  The other model parameters are $U^{\prime}=U-2J$, $t_2/t_1=0.5$,
  and the energies are in units of $t_1$.} \label{fig4}
\end{figure}

In Fig.~\ref{fig4}, we compare the phase diagrams relied on the NNN
hopping amplitude for the cases with different Hund's rule coupling
($J$): $J=U/8$, $J=U/4$, and $J=U/2$. As we know, Hund's coupling
is responsible for strong correlations in multiorbital systems.
The importance of $J$ in promoting orbital-selective physics
can be understood by recognizing that $J$ blocks orbital fluctuations
\cite{Georges-2013}. Just as expected, both $U_{c1}$ and $U_{c2}$ drop significantly when $J$ increases from $U/8$ (Fig.~\ref{fig4}(a))
to $J=U/2$ (Fig.~\ref{fig4}(c)), suggesting the enhancement of the
effective correlations. In order to make the computed results more clear,
in Fig.~\ref{fig4}(d) we plot the $J/U$ dependence of $\Delta$, which is
the difference of the critical values of the wide and narrow bands  ($\Delta=U_{c1}-U_{c2}$). The almost horizontal lines of $\Delta$ indicate that the region of the OSMP is almost unchange with the increasing $J$ in spite of the decreasing for the critical interactions for both wide and
narrow band. On the other hand, $\Delta$ for the system with NNN hopping
($K=1.0$) is near twice as large as that of the standard model with only
NN hopping, suggesting that the NNN hopping has an even  more obvious
effect on the OSMT than the Hund's rule coupling.

\section{Conclusions}
\label{conclusions}

We emphatically study the asymmetric effect introduced by the
next-nearest neighbor hopping on orbital-selective Mott
transition in the two-orbital Hubbard model. We find that the
asymmetric DOS introduced by the next-nearest neighbor hopping
strongly influences the energetics of the Mott gap. As a result,
the region of orbital-selective Mott phase increases significantly
with the increasing next-nearest neighbor hopping amplitude.
We also find that the orbital fluctuations are blocked with the
increasing of Hund's rule coupling, leading the critical values of
both narrow and wide band decrease manifestly.
However, the effect of the NNN hopping on OSMT is found to be more
significant than that of the Hund's coupling.

\section*{Acknowledgments}
The computational resources utilized in this research were provided by
Beijing Normal University high-performance scientific computing center.
The work was supported by the NSFC of China, under Grant No.
11174036 and 11474023, the National Basic Research Program of
China (Grant Nos. 2011CBA00108), and the Fundamental Research Funds for the
Central Universities.

\end{document}